# DNA Nucleobase Interaction Driven Electronic and Optical Fingerprints in Gallium Selenide Monolayer for DNA Sequencing Devices

Kuldeep Kumar[1] and Munish Sharma*[1]

[1]*Department of Physics, School of Basic and Applied Sciences, Maharaja Agrasen University, Baddi 174103, India.*

(February 9, 2021)

*Corresponding Author

Email: munishsharmahpu@live.com

**Abstract**

The interaction of DNA nucleobases with monolayer GaSe has been studied with in DFT framework using vdW functional. We found that nucleobases are physisorbed on the GaSe monolayer. The order of binding energy per atom is C > T > G > A. The room temperature recovery time estimated to be maximum of 113.88 μs for T+GaSe indicting reusability of the GaSe based devices. The modulation in the electronic structures of GaSe has been clearly captured within the simulated STM measurements. We also demonstrate quantum capacitance as a key parameter for sensing applications. Furthermore, in optical properties, electron energy loss (EEL) spectra show red shift in photon energy on nucleobase adsorption in UV region. The red shift is maximum of 0.92 eV for E⊥c and 0.50 eV for E∥c. In nutshell, GaSe monolayer exhibit anisotropic optical response in UV-region which can be highly beneficial for developing polarized optical sensors. Our results demonstrate that GaSe monolayer can be utilized to fabricate reusable DNA sequencing devices for biotechnology and medical science.

## 1. Introduction

The rapid pace of progress in two dimensional (2D) materials after graphene [1], transition metal di-chalcogenides (TMD's) [2, 3] and other 2D materials like hexagonal boron nitride [4], silicene [5], phosphorene [6], tellurene, arsenene etc. [7, 8] have attracted great deal of attention because these materials exhibit potential for various applications such as in solar cell, optoelectronic devices, bio-sensing [9-13] etc. In the field of bio-molecular analysis [14], disease diagnosis [15, 16] and forensic sciences [17], 2D nano-materials play a very important role because they have required high surface-to-volume ratio [18]. Various studies have been made to explore the applications of two dimensional materials like graphene [12, 19-21], graphene oxide (GO) [22, 23], metal oxide [24, 25] and $MoS_2$ [12, 13] based devices for nucleobase identification in long chain DNA. The lack of band gap in graphene [3, 26], high temperature operability in metal oxide and low electron and hole mobilities of $MoS_2$ monolayer at room temperature [27, 28] put these materials on back bench.

In the past few years, two dimensional gallium selenide (GaSe) from the family of post-transition-metal chalcogenides (PTMCs) has been reported in literature [29, 30]. The GaSe monolayers have two sub-layers of gallium (Ga) atoms sandwiched between two sub-layers of selenium (Se) atoms. In recent experiments P. Hu *et. al.* [29] and D. J. Late *et. al.* [30] synthesized few-layer GaSe nano-sheet using micromechanical cleavage technique. A 2D GaSe sheets can also be obtained by vapor-phase mass transport [31], vdW epitaxy [32], molecular beam epitaxy [33], and pulse laser deposition [34] methods. The two dimensional GaSe exhibits field-effect differential mobility of 0.6 $cm^2$ $V^{-1}$ $s^{-1}$. Also, the response time of 0.02 s, responsivity of 2.8 $AW^{-1}$ and external quantum efficiency of 1367% which make it useful for high performance nanoscale photo-detectors [29]. It has good current ON/OFF ratio (= $10^5$ ) and low dark current which make it suitable candidate for fabrication of nano-electronic devices [30] and transparent conductive support for studying biological molecules.

The investigation of essential bimolecular substances (amino acids and nucleobase) in human body is very important to extract and sequence the genetic information. Inspired from the superior properties of GaSe and its suitability for bio-sensing applications we here, devote our efforts to investigate potential of GaSe for the next generation DNA sequencing devices. TMD's have been investigated with similar motive but, they have shown its own limitations for distinguishing the four DNA nucleobases using electronic and optical methods [12, 13]. The gallium selenide (GaSe) from the family of post-transition-metal chalcogenides (PTMCs) represents an interesting alternative to TMD's which may be advantageous over present TMD's based bio-sensing techniques. Here, we have employed first principle approach to study the interaction of DNA nucleobase (viz. Adenine (A), Guanine (G), Cytosine (C) and Thymine (T)) with GaSe monolayer. Here, we demonstrate that it is possible to identify the nucleobase type by optical method and quantum capacitance.

## 2. Simulation Details

In this paper, DFT based calculations were performed within SIESTA code [35, 36]. The generalized gradient approximation (GGA) with Perdew-Burke-Ernzerhof (PBE) exchange function [37] was used to account the exchange and correlation energies. Here, the van-der-Waals interaction between nucleobase and GaSe was taken care by Grimme parameters [38]. Well tested Troullier Martin, norm conserving, relativistic pseudo-potential were used to treat electron-ion interactions. The double-zeta polarized (DZP) numerical atomic orbital basis set with the confinement energy 30 meV has been used to expand the Kohn-Sham orbitals. A mesh cutoff energy of 350 Ry has been used for reciprocal space expansion of the total charge density.

The unit cell of GaSe monolayer with 4 atoms was fully relaxed and a 5×5×1 supercell containing 100 atoms is used in calculations with vacuum of 16.94 Å along z-axis to minimize the interaction between the supercell images. A 3×3×1 k-points Monkhorst-Pack is used to sample Brillouin zone. All the geometries were relaxed until atomic forces minimized to 0.04 eV/Å. The energy dependence of nucleobase was check by rotating the nucleobase by $90^0$ in the steps of $10^0$. To calculate optical properties an optical mesh of 33×33×3 with 0.2 eV optical broadening is used.

## 3. Results and Discussion

### 3.1 Electronic Properties

The optimized structure of GaSe monolayer has been depicted in figure 1. It is found that in each case, nucleobases molecule occupied parallel configuration with GaSe monolayer which is in agreement with previously reported studies for MoS$_2$ [12, 13] and graphene [12, 39]. The sensitivity of GaSe monolayer has been gauged with binding energy per atom as

$$E_b = \frac{\{E_{GaSe+nucleobase} - (E_{GaSe} + E_{Nucleobase})\}}{n}$$

where, $E_{GaSe+nucleobase}$ denotes total energy of GaSe with absorbed nucleobase, $E_{GaSe}$ and $E_{Nucleobase}$ denotes energy of GaSe monolayer and isolated nucleobase respectively and $n$ denotes number of atoms in nucleobase. Calculated binding energy per atom ($E_b$) and optimum height (d) with respect to GaSe monolayer of composite system have been tabulated in table 1.

**Table 1.** Calculated binding energy per atom ($E_b$) and optimum height (d) of nucleobase above GaSe surface.

| System | $E_b$ (meV) | d (Å) |
|---|---|---|
| A+GaSe | -17.93 | 3.08 |
| C+GaSe | -36.77 | 3.29 |
| G+GaSe | -22.44 | 3.17 |
| T+GaSe | -31.99 | 3.19 |

Here, optimum height (d) is defined as the vertical distance between lowest C atom of nucleobase and the top Se atom of the GaSe monolayer. The binding energy per atom is maximum -36.77 meV for cytosine and minimum -17.93 meV for adenine. The order of binding energies per atom of nucleobase with GaSe monolayer has been found as C > T > G > A. A similar observation for maximum binding of cytosine have also been observed for silicene [10] and MoS$_2$ [13] while tellurene exhibit maximum binding energy per atom for guanine [9]. Furthermore, binding energy per atom varies between 5-19 meV which is smaller than the silicene [10] but higher than that for MoS$_2$ and tellurene [9, 12, 13]. Our calculated binding energy per atom values indicates that nucleobases are physisorbed on GaSe monolayer. The weak physisorption indicate expected less recovery time of nucleobase. The room temperature (at 300 K) recovery time estimated using transition state theory [40] show maximum recovery time of 113.88 μs for T+GaSe (Supporting Information). We find that with increase in temperature recovery time of respective nucleobase decreases (table S1). The smaller recovery time suggests GaSe as potential candidate for fabrication of reusable biosensors.

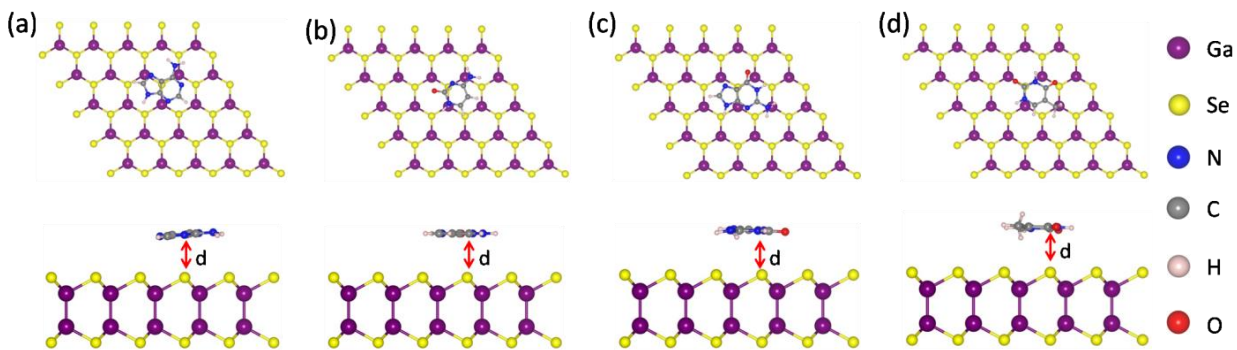

**Figure 1.** Top and side views of optimized geometries of nucleobases adsorbed on GaSe monolayer (a) A+GaSe (b) C+GaSe (c) G+GaSe and (d) T+GaSe.

To get an insight about nucleobase interaction with GaSe monolayer; charge density difference has been analyzed. The calculated charge density difference for nucleobase adsorbed GaSe monolayer has been presented in figure 2. The charge density difference has been calculated as $\Delta\rho = \rho_{GaSe+nucleobase} - (\rho_{GaSe} + \rho_{Nucleobase})$, where $\rho_{GaSe+nucleobase}$ represents total charge density of composite system, $\rho_{GaSe}$ and $\rho_{Nucleobase}$ represents charge density of pristine GaSe and isolated nucleobase respectively. It can be seen clearly from figure 2 that the charge redistribution occurs mostly on top Se atoms at the interface region due to proximity of nucleobase to Se atoms. The charge redistribution in adenine is lower than that of guanine which may be due to absence of oxygen-mediated interaction between adenine and GaSe. For interaction of guanine, polarization is stronger around nucleobase as well as between nucleobase and GaSe surface. It might be due to availability of oxygen and small optimum height from GaSe surface as compare to thymine and cytosine. The presence of two oxygen atoms in thymine reduces their significant simultaneous interactions with selenium atom due to geometric reasons. In addition, if we talk about contribution of different atoms then it is found that for the adsorption of adenine, cytosine, guanine and thymine mostly charge depletion and accumulation occurs between selenium atom of GaSe and oxygen atom, C-H and N-H of nucleobase. Overall on adsorption of all four nucleobases there is interaction of nucleobase with GaSe monolayer which can be characterized as weak and non-covalent. This weak interaction and physisorption is also supported by values of binding energy per atom which are less than 50 meV (table 1).

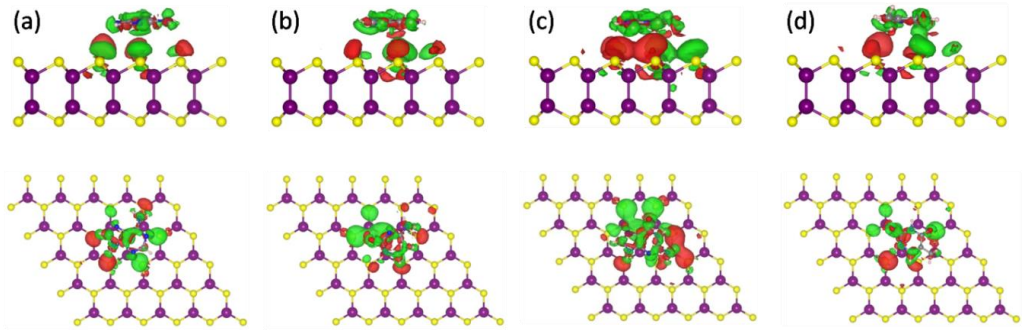

**Figure 2.** Side and top views of charge density difference profiles for (a) Adenine, (b) Cytosine, (c) Guanine and (d) Thymine adsorbed onto GaSe monolayer. Red and green color depicts charge accumulation and depletion respectively. Isosurface value is set at $23\times10^{-5}$ e/Å$^3$.

Furthermore, the electronic band structures have been calculated along Γ-M-K-Γ high symmetry point on the Brillouin zone for nucleobase+GaSe systems in equilibrium configuration which are shown in figure 3. The obtained band gap has been tabulated in table 2. The pristine GaSe show an indirect band gap of 1.87 eV, which is close agreement with previous reported DFT value of 1.83 eV [41] whereas experimental reported value is 2.10 eV [30]. The underestimation of experiential band gap is due to well-known artifact of exchange correlation functional with in the DFT. A deformation in CBM has been found for nucleobase+GaSe composite system as compared to pristine monolayer. The CBM deform by 2.04, 3.00, 3.16 and 16.55% for T, C, A & G complex respectively. Note that nucleobase modify the valence band in the vicinity of Fermi level resulting to decrease in effective band gap of GaSe monolayer. It has been found that maximum decrease in band gap occurs for guanine (18.50%) and minimum for thymine complex (1.12%). The order of band gap on adsorption is T > C > A > G.

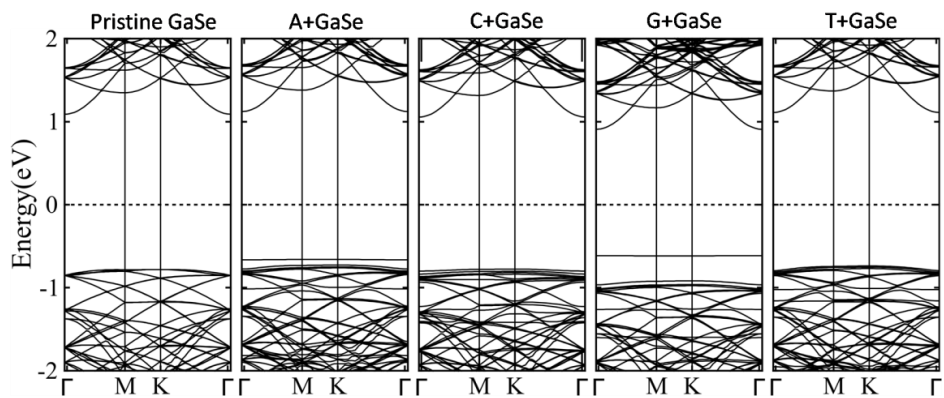

**Figure 3.** Electronic band structure for pristine and nucleobase+GaSe monolayer. Fermi level is set at 0 eV.

Furthermore, calculated work function ($\Phi$) of the pristine GaSe and nucleobase adsorbed GaSe has been tabulated in table 2. Work function has been computed as $\Phi = E_{vac} - E_f$, where $E_{vac}$ and $E_f$ represents the vacuum potential and Fermi energy respectively. The calculated work function of 4.96 eV for pristine GaSe monolayer is nearly close to the previous reported value of 5.59 eV by M. Yagmurcukardes *et al*. [42]. A modulation in work function occurs due to interactions of nucleobase with GaSe monolayer. The adenine and thymine increases the work function by 60 meV and 50 meV while cytosine and guanine decreases work function value by 20 meV and 150 meV respectively. The change in work function is attributed to the modulation in chemical potential due to nucleobase interaction.

**Table 2.** Calculated energy gap ($E_g$) and work function ($\Phi$) of pristine GaSe and nucleobase adsorbed GaSe.

| System | $E_g$ (eV) | $\Phi$ (eV) |
|---|---|---|
| GaSe | 1.87, 1.83[DFT], 2.10[Exp] | 4.96 |
| A+GaSe | 1.78 | 5.02 |
| C+GaSe | 1.83 | 4.94 |
| G+GaSe | 1.52 | 4.81 |
| T+GaSe | 1.85 | 5.01 |

[DFT]Ref[41], [Exp]Ref[30]

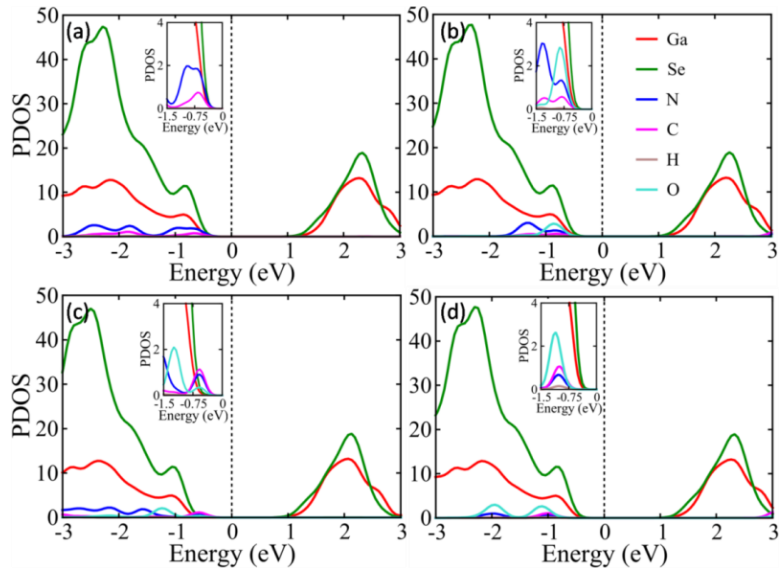

**Figure 4.** Atom projected density of states for (a) A+GaSe, (b) C+GaSe, (c) G+GaSe and (d) T+GaSe. The magnified PDOS for below the Fermi level (VB region) is presented in inset(s).

To confirm the molecular energy states in valence band, we have calculated atom projected density of states for nucleobase+GaSe system which is presented in figure 4. The presence of finite density of states at -0.5 eV for guanine complex and between -0.5 to -1.0 eV for other cases confirms molecular energy states in VB around Fermi level. The N and C DOS are more pronounced for adenine complex while O and N show contribution in VB for cytosine, guanine and thymine complex.

The Scanning tunneling microscopy (STM) images are helpful in understanding the structural and electronic information. The STM topographical images obtained within Tersoff and Hamann formalism [43, 44] for nucleobase adsorbed GaSe monolayer has been presented in figure 5. The simulated STM like measurements have been obtained at biasing V=+1.0V between sample and tip. The larger bright spots in the STM topographs originate from the nucleobase molecules due to presence of occupied energy states of nucleobase in valence band region. As the Se atoms contribute higher to DOS as compared to the Ga atoms, Se atoms expected to produce bright spots and Ga atoms less bright spots.

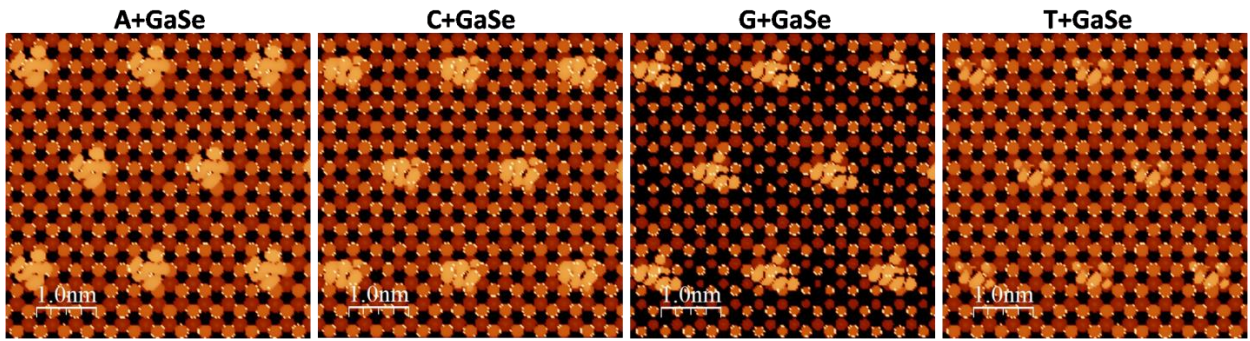

**Figure 5.** Simulated STM topographs for nucleobase+GaSe monolayer system for valence band at biasing of V=+1.0V between sample and tip.

## 3.2 Quantum Capacitance

Furthermore, quantum capacitance has been demonstrated as a key parameter in 2D sheets for gas-sensing application by A.H. Pourasl *et al.* [45]. The quantum capacitance ($C_Q$) can be defined as $C_Q = dQ/dV_a$, where $dQ$ is the variations of charge density and $dV_a$ is the local potential [46]. If electrochemical potential shifted by $eV_a$, then the excess charge density ($\Delta Q$) can be written as

$$\Delta Q = \int_{-\infty}^{+\infty} D(E)[f(E) - f(E - eV_a)]d(E)$$

where $D(E)$ is density of states, $e$ is the elementary charge ($1.6 \times 10^{-19}$C), $E$ is the relative energy with respect to Fermi energy $E_f$ and $f(E)$ is the Fermi-Dirac distribution function. For two dimensional materials, the analytical expression of quantum capacitance can be expressed as

$$C_Q = e^2 \int_{-\infty}^{+\infty} D(E) F_T(E - eV_a)d(E)$$

where $F_T(E)$ is the thermal broadening function and can be measured as

$$F_T(E) = (4k_B T)^{-1} sech^2(E/2k_B T)$$

where $k_B$ is Boltzmann constant and T is temperature which is 300K in our calculations.

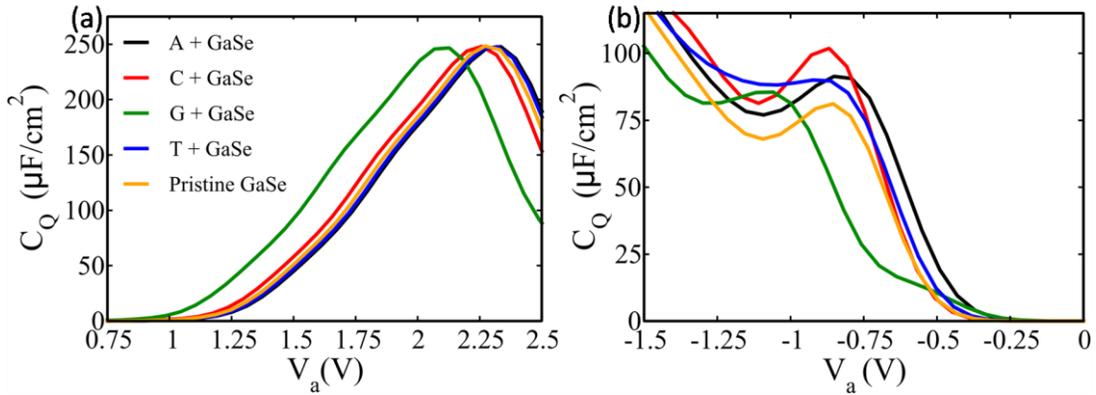

**Figure 6.** Quantum capacitance $C_Q$ for pristine and nucleobase adsorbed GaSe for (a) positive and (b) negative local electrode potential ($V_a$).

The effect of nucleobase adsorption on the quantum capacitance of GaSe is shown in figure 6. A considerable charge transfer at the nucleotide-surface interface results in modulation in quantum capacitance. At positive local potential $V_a$= +1.25 V, quantum capacitance is maximum for guanine and minimum for adenine (table 3). Order of quantum capacitance for this potential value up to +2.0 V is G > C > Pristine GaSe > T >A. However, at negative local potential $V_a$= -0.75 V, it just reverses i.e. maximum for adenine and minimum for guanine adsorbed system. For this local potential order of quantum capacitance is A > C > T > Pristine GaSe > G. Note that quantum capacitance changes significantly at lower value of negative local potential as compare to positive local potential. Furthermore, from these observations it is clear that the electronic states near the valence band are modulated and there is formation of impurities states in this region. Presence of these impurities states result in modulation of career concentration which leads to change in quantum capacitance.

**Table 3.** Quantum capacitance $C_Q$ for pristine and nucleobase+GaSe system at local potential +1.25 V and -0.75 V.

| System | $C_Q$ (μF/cm²) at | |
| --- | --- | --- |
|  | $V_a = +1.25$ V | $V_a = -0.75$ V |
| GaSe | 10.41 | 67.69 |
| A + GaSe | 7.92 | 86.03 |
| C + GaSe | 14.15 | 79.68 |
| G + GaSe | 37.89 | 27.94 |
| T + GaSe | 8.89 | 76.16 |

### 3.3 Optical Properties

The two dimensional materials have shown their capabilities for bio-molecular optical sensing applications [47]. The imaginary part ($\varepsilon_2$) of dielectric function and electron energy loss spectra for pristine and composite GaSe monolayer have been calculated for electric field vector perpendicular to c- axis (in-plane) and parallel to c- axis (out-of-plane) as shown in figure 7. The peak positions in imaginary part and energy loss spectra of pristine and composite GaSe monolayer have been tabulated in table 4.

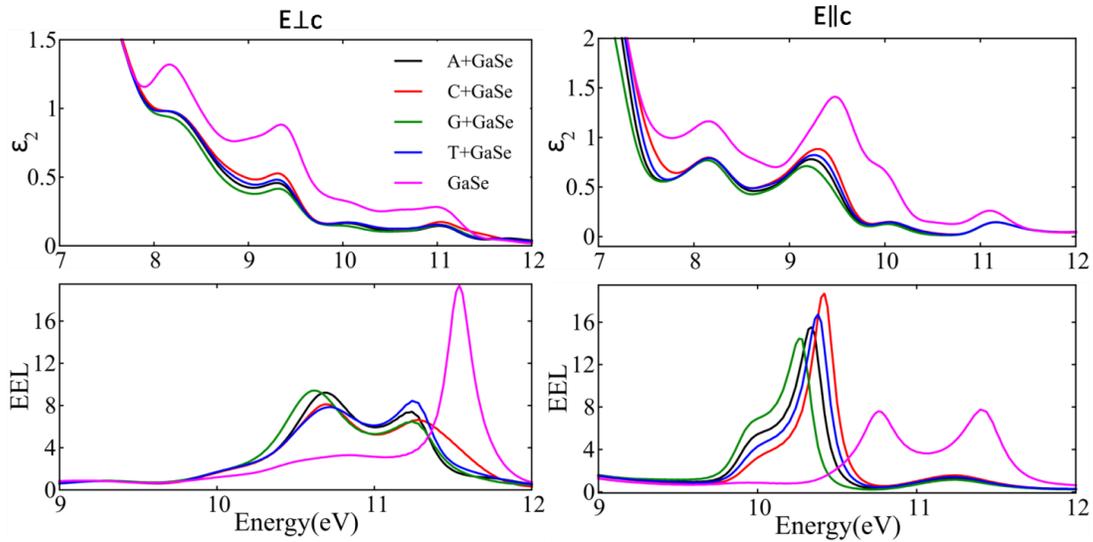

**Figure 7.** Imaginary part ($\varepsilon_2$) of dielectric function and electron energy loss (EEL) spectra for pristine and nucleobase adsorbed GaSe monolayer.

**Table 4.** Peak positions in imaginary part ($\varepsilon_2$) and electron energy loss (EEL) spectra of pristine and nucleobase adsorbed GaSe when E⊥c and E∥c.

| System | $\varepsilon_2$ (eV) | | EEL (eV) | |
|---|---|---|---|---|
| | E⊥c | E∥c | E⊥c | E∥c |
| GaSe | 9.34 | 9.48 | 11.54 | 10.76 |
| A+GaSe | 9.30 | 9.24 | 10.68 | 10.34 |
| C+GaSe | 9.30 | 9.31 | 10.69 | 10.42 |
| G+GaSe | 9.30 | 9.19 | 10.62 | 10.26 |
| T+GaSe | 9.30 | 9.26 | 10.71 | 10.38 |

For pristine GaSe monolayer the absorption peak ($\varepsilon_2$) appears at 4.2 eV, which is in close agreement with a previous study [48]. It is clear from figure S1 that ($\varepsilon_2$) for nucleobase+GaSe is identical to those of pure GaSe monolayer up to photon frequency of 7.5 eV for both polarizations (E⊥c and E∥c). A considerable modulation in $\varepsilon_2$ occurs in UV region (7.5-12eV). The peaks in UV region get red shifted for GaSe complex as compared to pristine GaSe monolayer. In out-of-plane polarization the maximum red shift of 0.29 eV is observed for the guanine and minimum of 0.17 eV for cytosine case. Notice that the shift is constant for in-plane polarization (E⊥c).

Furthermore, the electron energy loss (EEL) spectra has been calculated using the expression [49]:

$$\text{Im}\left\{\frac{-1}{\varepsilon(\omega)}\right\} = \frac{\varepsilon_2(\omega)}{\varepsilon_1^2(\omega) + \varepsilon_2^2(\omega)}$$

The electron energy loss (EEL) spectra correspond on collective excitations of electrons. For pristine GaSe monolayer a sharp peak is observed at 11.54 eV for E⊥c and relatively broader peak at 10.76 eV for E∥c. The peaks in EEL spectra show red shifted of 0.83 eV to 0.92 eV for nucleobase+GaSe as compared to pristine GaSe monolayer. The EEL spectra get red shifted by an amount of 0.83 eV for thymine and 0.92 eV for guanine complex for E⊥c. Where as in out-of-plane polarization peaks around 10.76 eV are red shifted by 0.34 eV for cytosine and 0.50 eV for guanine complex as compare to pristine GaSe and furthermore, this shift is non-uniform for all nucleobases. The red shift ranges from 0.34 eV to 0.50 eV. The real part of dielectric function for both the polarizations has been presented in figure S2. Frequency where $\varepsilon_1$ cuts the zero axis from the negative y-axis represents plasmon frequency ($\omega_p$). The calculated plasmon frequency for in-plane and out-of-plane polarization in different system has been tabulated in table 5. It can be seen form figure S2 that $\varepsilon_1$ cuts the zero axis from the negative y-axis at 11.52 eV and 10.77 eV respectively for E⊥c and E∥c. These values are consistent with peaks positions in EEL as given in table 4.

**Table 5.** Plasmon frequency ($\omega_p$) of pristine and nucleobase adsorbed GaSe for E⊥c and E∥c axis.

| System | $\omega_p$ (eV) | |
|---|---|---|
| | E⊥c | E∥c |
| GaSe | 11.52 | 10.77 |
| A + GaSe | 10.71 | 10.29 |
| C + GaSe | 10.71 | 10.39 |
| G + GaSe | 10.63 | 10.21 |
| T + GaSe | 10.74 | 10.34 |

## 4. Conclusions

In this paper, Density Functional Theory (DFT) within SIESTA code has been employed to study the interaction of nucleobases on GaSe monolayer. Following are the conclusions of our study:

- All the DNA nucleobases are physisorbed on GaSe monolayer and show room temperature maximum recovery time of 113.88 μs for T+GaSe.
- Order of binding energy per atom is C > T > G > A.
- A considerable polarization of GaSe monolayer occurs due to nucleobase adsorption.
- On adsorption of nucleobase electronic band gap decreases. Maximum decrease in band gap is for guanine + GaSe (18.50%) and minimum for cytosine + GaSe (1.12%) as compare to pristine GaSe.
- Some new features occur in density of states of GaSe due to adsorption of nucleobases which is also confirmed by simulated STM topographical images.

- There is modulation in work function of GaSe due to adsorption of nucleobases. Adsorption of adenine and thymine increases the work function while cytosine and guanine decreases the work function of GaSe.
- A considerable modulation in quantum capacitance occurs on adsorption of nucleobases which can be attributed to charge transfer at the nucleotide-surface interface.
- EEL of GaSe + nucleobase conjugate system is red shifted by 0.83 to 0.92 eV for E⊥c, where as it is red shifted by 0.34 to 0.50 eV for E∥c.

Therefore, our present study show that GaSe monolayer has a good potential in DNA sequencing and it can be a promising candidate for reusable biosensors to detect DNA nucleobases.

## Conflict of Interest
There is no conflict of interest to declare.

## Acknowledgements
A valuable discussions with Ritika Rani, Manoj Kumar, Vivek and Jyoti Rai are highly acknowledged.